\documentstyle[twocolumn,aps,epsf]{revtex}
\begin{document}
\draft
\title{Dynamics of a Strongly Damped Two-Level System: Beyond the DBGA
\thanks{Preprint no.: IMSc/94-95/65}}

\author{Tabish Qureshi\cite{email}}

\address{Institute of Mathematical Sciences,\\ C.I.T. Campus, Taramani,
Madras-600113, INDIA}
\maketitle
\begin{abstract}
Dynamics of a dissipative two-level system is studied using
quantum relaxation theory. This calculation for the first time
goes beyond the commonly used dilute bounce gas approximation
(DBGA), even for strong damping. The new results obtained here
deviate from the DBGA results at low temperatures, however,
the DBGA form is recovered at high temperatures. The results
in the parameter regime $ 1/2<\alpha  <1$, where the model has
connection with the Kondo Hamiltonian, are of particular significance.
In this regime, the spin shows a cross-over to a slower exponential
relaxation at intermediate times, which is roughly half the
relaxation rate at short times, as also observed in Quantum
Monte-Carlo simulation of the model. The asymptotic behavior of the
spin in the Kondo regime is in agreement with the exact conformal
field theory results for the Kondo model. A connection of the
dissipative dynamics of the two-level system with the quantum
Zeno effect is also presented.
\end{abstract}


\section{introduction}

For the task of describing the interesting phenomenon of dissipation
in quantum systems, two-level systems have served as a simple
and tractable prototype. Over the years, several experimental
situations have been found which can be accurately described
in terms of damped two-level systems. Superconducting Quantum
Interference Devices (SQUIDs) for example, are systems where
quantum coherence can be studied at a macroscopic level and
description in terms of a two-level system is reasonably good.
In metallic glasses, where atoms are quenched in random positions,
certain atoms can end up finding themselves in a situation where
two close by positions are energetically equivalent \cite{Black}.
So it is easiest for them to quantum mechanically tunnel between
the two positions. Such a system, at very low temperatures, behaves
like a two-level system. The sea of conduction electrons however
impedes this motion. What one ends up with is a dissipative two-level
system. Similar situation exists for light interstitial particles
in metals, which has become an interesting field of its own
and goes by the name of ``hydrogen in metals'' \cite{Wipf}. An
example from the realm of quantum optics is an atom in a radiation
field. If one is interested in the radiative properties of just one
excited state of the atom, one can imagine it to be a two-level
system coupled to a dissipative ``bath'' of photons. A very
well studied problem in condensed matter physics, that of a
static spin 1/2 magnetic impurity in a metal (the so-called
Kondo problem) \cite{Kondo}, can also be thought of as a two-level
system coupled to the conduction electrons. Later in this paper, we
shall see some detailed application of the results to this problem.
Recent interest in ``quantum computers'',  has led to the introduction
of some interesting models to address the problem of quantum coherence
\cite{Unruh,Zurek}. These too can be recast in the language of dissipative
two-state systems. Lately, in the active field of high temperature
superconductivity, dissipative two-level systems have been recalled
in the context of c-axis transport in the normal state of some
high temperature superconductors \cite{Chakravarty,Clarke}.

A dissipative two-level system can be generally described by
the so-called spin-boson Hamiltonian \cite{Leggett}. It consists of a
psuedo-spin $\sigma$, depicting the two-level system and a set of
independent harmonic oscillators. The coordinates of the harmonic
oscillators are {\it linearly } coupled to the z-component of the
pseudo-spin $\sigma_{z}$. The Hamiltonian can be expressed as
\begin{equation}
 H = { 1\over 2} \hbar \Delta \sigma_{x }+ \sum^{ \infty }_{ j =
0 }\biggl( { p_{j}^{2}\over 2m} + { 1\over 2} m_{j}\omega_{j}^{2
}(x_{j }-{c_{j}\over m_{j}\omega_{j}^{2}}\sigma_{z })^{2 }\biggr) ,
\end{equation}
where $ \Delta $ is like a tunneling matrix element if one
has a particle in a double-well kind of a system in mind, $ x
 _{j }s $ and $ p_{j }s $ are the coordinates
and momenta of the harmonic oscillators, and $ c_{j }s
$ are their respective coupling constants to the two-level system,
i.e., the pseudo-spin. Looking at the Hamiltonian in (1) one
would notice that the (pseudo-)spin by virtue of being in a
state with $ \sigma_{z }= $ +1 or -1, shifts the centers
of the harmonic oscillators to the left or right. This results
in a dissipative drag on the spin. It turns out that in order
to analyse the influence of the ``heat-bath'' consisting of
harmonic oscillators on the spin $ \sigma , $ one need not demand
the individual knowledge of $c_{j}$s, $m_{j}$s and $\omega_{j}$s.
One only requires them in a particular
combination as they appear in the so-called spectral density
function :
\begin{equation}
 J(\omega ) = \sum^{ \infty }_{ j = 0 }{ c_{j}^{2}\over
m_{j}\omega_{j}} \delta (\omega -\omega_{j }).
\end{equation}
The spectral density function contains all the information needed
to specify the dissipative dynamics of the spin $ \sigma .
$ In order to describe the dissipative behavior and introduce
irreveresibilty one assumes a {\it continuous } spectrum
of frequencies in the bath. The most common form of the spectral
density is the ``Ohmic'' form where $ J(\omega ) $ is linear
in $ \omega $ for small frequencies:
\begin{equation}
 J(\omega ) = \alpha \omega e^{-\omega /\omega_{c} },
\end{equation}
where $ \alpha $ is a dimensionless coupling constant parametrizing
the strength of the interaction of the spin with the heat-bath,
and $ \omega_{c }$ is a large cutoff frequency. The
Ohmic form of the bath describes a variety of systems for various
values of $ \alpha . $ For example, for $ 0\le \alpha \le $ 1/2,
the spin-boson Hamiltonian describes the low-temperature tunneling
dynamics of a positively charged particle in a double-well potential,
in the presence of conduction electrons \cite{Wipf}. The harmonic
oscillators in this case represent charge density excitations of the
electron gas, which behave like bosons. For the case $1/2<\alpha <1$,
the spin-boson Hamiltonian mimics the Kondo model. The harmonic
oscillators here play the role of spin density excitations of
the electron gas, and the pseudo-spin symbolizes the magnetic
impurity spin. For most purposes, the quantity of interest is
the symmetrized spin correlation function
\begin{equation}
 C(t) = { 1\over 2} <[\sigma_{z }(0)\sigma_{z }(t) + \sigma_{z}
(t)\sigma_{z }(0)]> ,
\end{equation}
where the angular brackets denote thermal average, and the Heisenberg
time evolution of $ \sigma_{z }$ is dictated by the Hamiltonian (1).

The spin-boson model is a very old one, but for reasons mentioned
earlier, it still attracts considerable research attention.
Its dynamics is what has been of most interest to researchers.
It was first attacked by Leggett and collaborators using the
path-integral formalism of Feynman and Vernon \cite{Leggett}. Using
the path-integral method, it possible to {\it exactly} integrate
over the bath degrees of freedom to obtain a reduced dynamics of
the pseudo-spin, at least formally. But in order to obtain closed
expressions for the reduced dynamics, one has to make certain
approximations. The authors employed the so-called dilute bounce
gas approximation (DBGA) to the functional integral expression.
The results thus obtained turned out to be very good for a wide
range of values of the parameter $ \alpha $ and temperature.
The functional integral analysis of the spin-boson Hamiltonian
based on the instanton technique was considered by many as quite
elaborate and formidable. This led to some simpler derivations
of the DBGA results using second order Born approximation
\cite{Aslangul,Dekker} and resolvent expansion techinque \cite{SD}.
These calculations demonstrated the precise manner in which the DBGA
is connected to conventional perturbative techniques.

The DBGA was widely applauded for the simplicity of the results
obtained and the accuracy with which they described the dynamics.
For the case $ 0\le \alpha <1/2$, the spin shows weakly damped
coherent oscillations at low temperatures. The physical picture
is the following. The spin evolves quantum mechanically and
is decohered by the effect of the heat-bath over a time which
is much longer than $ \Delta^{-1 }$. So the spin
has time to evolve quantum mechanically before its coherence
is destroyed. As the temperature is raised, the excitations
of the heat-bath increase in strength, and at a particular temperature,
the time over which coherence is destroyed is much smaller than
$ \Delta^{-1 }. $ Here the coherent oscillations
of the spin disappear completely, and what is left is an incoherent
relaxation. Incoherence sets in faster if $ \alpha $ is large.
When $ \alpha $ becomes equal to 1/2, coherence is destroyed
completely at all temperatures. For $ 1/2<\alpha <1$, the
spin mimics a Kondo impurity spin and shows relaxation without
any coherent behavior. In the region $ 0\le \alpha <1/2$,
the DBGA is very good at not very low temperatures and at not
very long times. In the incoherent relaxation behaviour, the
spin decays with a rate which follows a power-law with temperature.
At very low temperatures DBGA breaks down unless the coulping
$ \alpha $ is very weak. There have been some calculations
which have gone beyond the DBGA for weak coupling \cite{Weiss,sdtq}.
It has been demonstrated that for an unbiased two-level system, for
weak coupling, those calculations essentially yield the DBGA result.
Thus it seems that the so-called ``inter-bounce'' interactions,
at low temperature, vanish as $ \alpha^{2 }, $ which
means that for strong coupling, inter-bounce interactions have
to be taken into account. This territory has been explored in
the present investigation. In the Kondo regime too, the DBGA
is quite good at not very long times, or small frequencies.
In fact, DBGA yielded new analytical results for the dynamics
of the Kondo spin, which reproduce many old perturbative results
for the Kondo problem, in various limits, and generate very
good fits to spectroscopic data in dilute magnetic alloys
\cite{tqkondo}. Inspite of this success, DBGA fails badly at long
times, and the question regarding the low temperature, long times
dynamics of the Kondo spin remains open. In fact, behavior of C(t)
in the regime $T=0$, $1/2<\alpha \le 1$ is considered a currently
unresolved problem (see section IV E of \cite{Leggett}).

In the present study, I put forward a calculation based on the
resolvent expansion technique. The results achieved, detail
the dynamics of the spin at low, as well as high temperatures,
for all relevant values of $ \alpha$. I shall present some
new results in the difficult regime mentioned in the last para.
The asymptotic behavior of the Kondo spin seen here agrees with
the {\it exact} asymptotic results derived by Affleck and
Ludwig for the Kondo problem using conformal field theory
\cite{Affleck}. This serves as a stringent test of the approximation
employed here.

\section{``non-perturbative'' expansion}

I begin by making a unitary transformation on the
Hamiltonian which pulls back the centers of the shifted oscillators
by the amount $ c_{j}\sigma_{z }. $ The unitary operator
for the purpose is given by $ S \equiv e^{-2\sigma_{z}\xi }$ where
$\xi =\sum c_{j}^{2}p_{j}$. The operation $SHS^{-1}$ on (1) leads to
\begin{equation}
 H' = { 1\over 2} \hbar \Delta (\sigma_{- }e^{-
\xi }+ \sigma_{+ }e^{\xi })
+ \sum^{ \infty }_{ j = 0 }\biggl( { p_{j}^{2}\over 2m} + { 1\over
2} m_{j}\omega_{j}^{2 }x_{j}^{2 }\biggr) .
\end{equation}
Now I expand the exponential factors in the first term in (1)
in a Taylor series and write them as
\begin{equation}
 e^{\pm \xi }= 1 + \sum^{ \infty }_{ n=1 }{ (\pm \xi )^{n}\over
n!} .
\end{equation}
Substituting (6) in (5) and regrouping terms I obtain
\begin{eqnarray}
 H &=& { 1\over 2} \hbar \Delta \sigma_{x }+ { 1\over 2} \hbar \Delta
\bigl\{\sigma_{x}\sum^{ \infty }_{ n=1 }{ \xi^{2n}\over (2n)!}
+ i\sigma_{y}\sum^{ \infty }_{ n=1 }{ \xi^{2n-1}\over
(2n-1)!} \bigr\}\nonumber\\
&& + \sum^{ \infty }_{ j = 0 }\biggl( { p_{j}^{2}\over
2m} + { 1\over 2} m_{j}\omega_{j}^{2 }x_{j}^{2}\biggr) .
\end{eqnarray}
Finally, I rotate my spin space basis about the y-axis, by $ \pi/2$
to arrive at
\begin{eqnarray}
 H'' &=& { 1\over 2} \hbar \Delta \sigma_{z }+ { 1\over 2} \hbar
\Delta\bigl\{\sigma_{z}\sum^{ \infty }_{ n=1 }{ \xi^{2n}\over (2n)!}
+ i\sigma_{y}\sum^{ \infty }_{ n=1 }{ \xi^{2n-1}\over
(2n-1)!} \bigr\}\nonumber\\
&& + \sum^{ \infty }_{ j = 0 }\biggl( { p_{j}^{2}\over
2m} + { 1\over 2} m_{j}\omega_{j}^{2 }x_{j}^{2}\biggr) .
\end{eqnarray}
Notice that in the new basis while $ \sigma_{x }$ has
become entangled with the bath operators, $ \sigma_{z}$ is
just rotated to $ -\sigma_{x }. $ Consequently,
for evaluating $C(t)$ one just has to replace $ \sigma_{z}$ by
$ -\sigma_{x }, $ and $ H $ by $H''$. Now
the cleverest way to do perturbation theory would be to locate
terms depending either only on the spin coordinates or only
on the bath coordinates, and treat them exactly. Following this
idea I split the Hamiltonian in (8) into a Hamiltonian for
the spin $ H_{S }, $ Hamiltonian for the bath $ H_{B }$, and
the Hamiltonian for the spin bath interaction $ H_{I }$, given by
\begin{eqnarray}
 H_{S} &=& { 1\over 2} \hbar \Delta \sigma_{z }, \nonumber\\
 H_{B} &=& \sum^{ \infty }_{ j = 0 }\biggl( { p_{j}^{2}\over 2m} +
{ 1\over 2} m_{j}\omega_{j}^{2 }x_{j}^{2}\biggr) , \nonumber\\
 H_{I} &=& { 1\over 2} \hbar \Delta \bigl\{\sigma_{z}
\sum^{\infty}_{ n=1 }{ \xi^{2n}\over (2n)!} + i\sigma_{y}
\sum^{\infty}_{ n=1 }{ \xi^{2n-1}\over (2n-1)!} \bigr\} .
\end{eqnarray}
It is now possible to treat $ H_{I }$ as a perturbation,
which does {\it not } amount to a weak coupling between the
spin and the bath. One will notice that even if $ H_{I}$ is
treated to second order, the final expression will contain
$ \alpha $ to arbitrary order due to the infinite series in
$ \xi$. The more important point is that, in addition to
$ \alpha$, the final expression will also contain $\Delta$ to
arbitrary order, due to the exact treatment of $ H_{S }$.
This is very important from the point of view of the Kondo problem,
because $ \Delta $ is proportional to $ J_{\perp }$, the
transverse part of the anisotropic exchange coupling in the
Kondo Hamiltonian \cite{Leggett}. This modus operandi will keep
both $ J_{\perp }$ and $ J_{\parallel }$ to all orders
and hence is expected to produce a much better perturbation
theory.

Instead of looking at the spin correlation function $C(t)$ itself,
it is convenient to handle its Laplace transform $\hat{C}(s)$:
\begin{equation}
 \hat{C}(s) = \int^{ \infty }_{ 0 }e^{-st }C(t) dt ,
\end{equation}
where $ s $ is the Laplace transform variable. In order to write
out the Laplace transform of (3) explicitly, the time evolution
of $ \sigma_{z }$ in the Heisenberg representation
can be expressed as $ exp({i\over\hbar} H''t)\sigma_{z}
exp(-{i\over\hbar} H''t)$. Introducing the Liouvillian associated
with $ H $ as $L(..)={ 1\over\hbar} [H,..]$, the Heisenberg evolution
of $ \sigma_{z}$ can be expressed in a more compact notation
$ e^{iLt}\sigma_{z }$. The ``superoperator'' $e^{iLt }$ completely
specifies the time evolution
of the system and is referred to as the time evolution operator,
denoted by  $U(t)$. Further, I assume that the initial density matrix
of the total system is factorized as $\rho_{S}\cdot \rho_{B}$. The
Laplace transformed correlation function then assumes the form
\begin{equation}
 \hat{C}(s) = Trace \biggl( \rho \bigl\{\sigma_{x }({ 1\over s-iL}
\sigma_{x}) + ({1\over s-iL} \sigma_{x}) \sigma_{x}\bigr\} \biggr),
\end{equation}
where the $ (s-iL)^{-1 }$ is the Laplace transform
of the time evolution operator. As the $ \sigma s $ do not depend
on the bath variables, one can perform a trace over the bath
variables to obtain a ``bath averaged'' time evolution operator
$ <\hat{U}(s)>_{B }, $ rather its Laplace transform. The tracing
of the bath variables can be formally performed by introducing
a projection operator $ P $ which when acting on an operator $A$
is defined by $PA \equiv Trace_{B }(\rho_{B }A)$.
All the information regarding the effect of the bath on the
spin is contained in the averaged time evolution. The major
task then is to evalute $ <\hat{U}(s)>_{B }$ in a suitable
approximation. The average time evolution operator can be shown to
satisfy the following integro-differential equation \cite{Dattagupta}
\begin{eqnarray}
 { d\over dt}<U(t)>_{B }&=& iL_{S }<U(t)>_{B} \nonumber\\
&&+ \int^{t}_{0}
[M(t-t )]_{av }<U(t )>_{B }dt ,
\end{eqnarray}
where $[M(t-t)]_{av }$ is called the memory function. The Laplace
transform of the memory function looks like
\begin{eqnarray}
[\hat{M}(s)]_{av }&=& P(iL_{I })(1-P)\nonumber\\
&&{ 1\over s - iL_{S }-
iL_{B}-(1-P)(iL_{I })(1-P)}\nonumber\\
&&(1-P)(iL_{I })P,
\end{eqnarray}
where $ L_{S}$, $L_{B}$ and $ L_{I}$ are the Liouvillians associated
with $ H_{S}$, $H_{B }$ and $ H_{I }, $ respectively.
In terms of the memory function, the average density operator
has deceptively compact form
\begin{equation}
 <\hat{U}(s)>_{B }= { 1\over s - iL_{S }- [\hat{M}(s)]_{av}} .
\end{equation}
The form is deceptive because $ <\hat{U}(s)>_{B }$ is still
{\it exact} and a complicated object to calculate. I shall
restrict myself to treating $ [\hat{M}(s)]_{av }$ to
second order in $ L_{I }, $ which is manageable. The corresponding
expression for the memory function takes up the subsequent form :
\begin{eqnarray}
 [\hat{M}(s)]_{av }&\approx& i(PL_{I }) + (PL_{I }){ 1\over s - iL_{S}}
(PL_{I }) \nonumber\\
&&+ P\bigl\{ L_{I}{ 1\over s - iL_{S }- iL_{B }} L_{I}\bigr\}.
\end{eqnarray}
The strategy is to first perform the trace over the bath states
in (15), which will give a $ 4\times 4 $ matrix in the spin
states. This matrix can then be plugged in the denominator of
(14). A $ 4\times 4 $ matrix inversion then yields the averaged
time evolution operator $ <\hat{U}(s)>_{B }$ which one requires.
The matrix elements of the last term in (15) can be written
in terms of correlation functions of certain bath operators,
which can be calculated using the properties of a bath of independent
harmonic oscillators alone. In the ensuing analysis the spin
states will be denoted by Greek indices, and the number states
of a set of harmonic oscillators by $ \mid n>$, $ \mid n'>$ etc.
We assume a canonical form of the density matrix at a temperature
$T$. The Laplace transformed spin correlation function can be
written in terms of the matrix elements of $ <\hat{U}(s)>_{B }$ :
\begin{eqnarray}
 \hat{C}(s) &=& { 1\over 2} \bigl\{ (+-\mid <\hat{U}(s)>_{B}\mid +-)
+ (+-\mid <\hat{U}(s)>_{B}\mid -+) \nonumber\\
&& + (-+\mid <\hat{U}(s)>_{B}\mid -+) + (-+\mid <\hat{U}(s)>_{B}\mid +-)
\bigr\},\nonumber\\
\end{eqnarray}
where $ \mid \mu \nu)$ denote Liouville ``states'' for the
spin, $ \mid \pm >$ being the eigenstates of $ \sigma_{z }.$ One
may also be interested in the average value of the spin
$\sigma_{z }$ assuming it started out at $t=0$ with
the value (say) $ \sigma_{z}=+1$. The Laplace transform
of $<\sigma_{z }(t)>$ is denoted in the literature by $\hat{P}(s)$. In
the present approximation scheme, $\hat{P}(s)$ turns out to be identical
to $\hat{C}(s)$. Another quantity of interest, especially for the Kondo
problem, is the imaginary part of the spin susceptibility
\begin{equation}
\chi''(\omega ) = {1\over 2\pi}\int^{\infty}_{-\infty}e^{-i\omega t}
<[\sigma_{z }(t),\sigma_{z}(0)]> dt,
\end{equation}
which can be rewritten in terms of a Laplace transform as
\begin{equation}
 \chi ''(\omega ) = {1\over\pi}Re\int^{ \infty}_{0}
e^{-st}<[\sigma_{z }(0),\sigma_{z }(t)]> dt.
\end{equation}
\begin{eqnarray}
 \chi '' (\omega ) &=& (<+\mid \rho_{S}\mid +> - <-\mid \rho_{S}\mid->)
\nonumber\\
&&\bigl\{ (+-\mid <\hat{U}(s)>_{B}\mid +-)
+ (+-\mid <\hat{U}(s)>_{B}\mid -+) \nonumber\\
&&+ (-+\mid <\hat{U}(s)>
_{B}\mid -+) + (-+\mid <\hat{U}(s)>_{B}\mid +-) \bigr\}.\nonumber\\
\end{eqnarray}
Some times a quantity, referred to as the spectral function,
is defined as $ S(\omega ) = \chi ''(\omega )/\omega$.
 This is a symmetric function of $ \omega . $ Now the stage is
set to compute the time evolution operator and put the corresponding
matrix elements in (16) and (19).

\section{Results}

I will skip the details of the algebra involved, which was handled
using Mathematica, and present the final form of the averaged
time evolution operator. Only the portion of the $ 4\times 4$ matrix for
$<\hat{U}(s)>_{B }$, within the space $\mid +-)$, $\mid -+)$ is displayed :
\begin{equation}
 <\hat{U}(s)>_{B }= { 1\over Det} \pmatrix{s - F(\omega ) + F(\omega
_{+ }) & F(\omega ) \cr F(\omega ) & s - F(\omega ) + F(\omega_{-})}
\end{equation}
where
\begin{eqnarray}
 Det &=& [s + F(\omega_{+ })][s + F(\omega_{- })]\nonumber\\
&& + F(\omega )
[z + { 1\over 2} \{F(\omega_{+}) + F(\omega_{-})\}].
\end{eqnarray}
In (20) and (21) $ \omega_{\pm }= \omega $
$ \pm \Delta$, and $F(\omega)$ is the Laplce transform (with $s=i\omega$)
of the following quantity,
\begin{eqnarray}
 F(t) &=& \Delta^{2}exp \biggl( -2\alpha \int^{\infty}_{0}{ 1\over
\omega } e^{-\omega /\omega c }[\coth(\hbar \beta \omega)\nonumber\\
&&\{1-cos(\omega t)\} + i\sin(\omega t)]d\omega \biggr) + c.c.
\end{eqnarray}
The correlation function (16) now looks like
\begin{equation}
 \hat{C}(s) = { 1\over s + F(\omega ) + [F(\omega_{+ })+F(\omega_{- })]/2
- { [F(\omega_{+ })-F(\omega_{- })]^{2 }/4
\over s + [F(\omega_{+ })+F(\omega_{- })]/2} } .
\end{equation}
The integral in (22) can be done for $t$, $ \hbar \beta \gg 1/\omega_{c}$,
so that its Laplace transform assumes the following structure
\begin{equation}
 F(\omega ) = { \Delta^{2}\over 2\omega_{c}} \biggl({ 2\pi \over\hbar
\beta \omega_{c}}\biggr)^{2\alpha-1 }{ \Gamma (1-2\alpha)
\Gamma (\alpha +i\omega \hbar \beta /2\pi )\over \Gamma (1-\alpha
+i\omega \hbar \beta /2\pi)} cos(\pi \alpha ),
\end{equation}
where $ \Gamma $ is Euler's Gamma function. Similarly,
$\chi ''(\omega )$ becomes
\begin{eqnarray}
&&\chi '' (\omega ) = {1\over\pi} Re \nonumber\\
&&{ \tanh(\hbar \beta \Delta /2)
[F(\omega_{+})-F(\omega_{-})]/2\over [s+F(\omega_{+})][s+F(\omega_{-})]
+ F(\omega )[s + { 1\over 2} {F(\omega_{+ })+F(\omega_{- })}]}.
\end{eqnarray}
The correlation function expressed by (23) describes the complete
dynamics of the spin in the spin-boson model. For T=0, one can
do better than this by taking the limit $ T\rightarrow 0 $ in (22)
itself. In this case we do not need to make any approximation,
and the exact form of $ F(\omega ) $ is given by
\begin{eqnarray}
 F(\omega ) &=& -i(\omega_{c}/2)(\Delta /\omega_{c })^{2 }
(\omega /\omega_{c })^{2\alpha -1 }e^{\omega /\omega c }
\Gamma (1-2\alpha,\omega /\omega_{c })\nonumber\\
&&-i(\omega_{c}/2)(\Delta /\omega_{c })^{2 }
(-\omega /\omega_{c })^{2\alpha -1 }\nonumber\\
&&e^{-\omega /\omega c }
\Gamma (1-2\alpha,-\omega /\omega_{c }),
\end{eqnarray}
where $ \Gamma $ (a,b) is Euler's incomplete Gamma function.
Interestingly, $ \omega_{c }$ can be completely ``scaled
out'' of (26), and hence the correlation function.

\section{discussion}

The equations (23) and (25) describe the dynamics of the spin
at all temperatures and all values of the coupling strength.
For the sake of comparison let us look at the corresponding
DBGA result, which has the following form :
\begin{equation}
 \hat{C}(s) = { 1\over s + 2F(\omega )} .
\end{equation}
The reader will notice that if the $ \Delta $ dependence in
the arguments of the Gamma functions in (23) is ignored, it
reduces the to the DBGA result (27). Let us discuss the different
parameter regimes in somewhat greater detail.
\epsfysize=2.9in
{\raggedright Figure 1 {\it
$Re\hat{C}(s)$ plotted against $\omega$ for $\omega_{c}
/\Delta=1000$, $\alpha=0.1$ and $k_{B}T/\hbar\Delta = 0.5$.
The solid line denotes the present calculation whereas the
dashed line denotes the DBGA result.}}
\bigskip

\subsection{The case $ 0\le \alpha <1/2$}

For this range of values of $\alpha$, the spin shows a coherent
evolution at low temperatures. This coherent evolution is damped
by the dissipative interaction. In this situation the spin-boson
model can describe the coherent tunneling behavior of a particle
in a double well potential, interacting with conduction electrons.
Figures 1 shows real part of $\hat{C}(s)$ which is related to the
structure factor for neutron scattering off the tunneling particle.
The correlation function shows two `inelastic' peaks which is a
signature of coherent evolution. The position of the peaks is clearly
shifted from the DBGA values.

As the temperature is raised the peaks broaden and shift towards
the origin. At a characteristic temperature the function assumes
the form of a single peak at $\omega =0$. The coherent behavior
is completely destroyed and the spin relaxes incoherently with
an exponential decay rate. Figure 2 shows the crossover as a
function of temperature. It is convenient to introduce a frequency
$ \Delta_{r }\equiv \Delta (2\pi /\hbar \beta \omega_{c })^{\alpha}
\sqrt{\Gamma (1-2\alpha)}$ which is approximately the effective tunnel
splitting within the DBGA. For $ T \gg \Delta_{r }$ the correlation
function $\hat{C}(s)$ can be approximated by
\begin{equation}
 \hat{C}(s) \approx { 1\over s + F(0) + [F(\Delta )+F(-\Delta )]/2 +
{ [F(\Delta)-F(-\Delta )]^{2 }/4\over [F(\Delta )+F(-\Delta )]/2} } .
\end{equation}
The expression in (28) depicts a spin relaxing with an exponential
decay rate which goes like $ T^{2\alpha-1 }$. The tunneling partical tends
to localize as temperature is increased. This is very counter-intuitive
but has been observed in the several experiments on two-level systems.
For still higher temperatures $ k_{B }T\gg \hbar \Delta , $ one can
write $ F(\Delta ) \approx F(-\Delta ) \approx F(0)$. For this case the
expression (28) reduces to the DBGA result(27).
\epsfysize=2.9in
{\raggedright Figure 2 {\it
$Re\hat{C}(s)$ plotted against $\omega$ for $\omega_{c}
/\Delta=1000$, $\alpha=0.1$ and various values of temperature
$\tau = k_{B}T/\hbar\Delta$. }}

\subsection{Relation to quantum Zeno effect}

One way to understand this decrease in relaxation rate is in
terms of the so-called quantum Zeno effect. This effect was
proposed by Mishra and Sudarshan \cite{Mishra} for a quantum
system on which a series of measurements are made. The limit
of `continuous' measurement results in a freezing of the `free'
dynamics of the system completely. In the problem at hand, the
heat-bath consisting of the harmonic oscillators is sensitive
to the position of the particle, the $ \sigma_{z }$ -state
of the spin in the problem at hand, as is obvious from the form
of the Hamiltonian (1). In some sense the heat bath monitors
the position of the particle in the double-well. By virtue of
its posessing infinite degrees of freedom, it is capable of
`collapsing' the wave function of the tunneling particle to
a particular well. With increasing temperature the measurement
becomes more and more efficient. In addition, the interaction
being present all the time amounts to a continuous measurement
of the particle being in one of the two wells. Thus the coherent
dynamics of the particle between the wells is destroyed and
it tends to `freeze' in one of the wells. Here one can easily
work out the `decoherence time scale' which has been a topic
of some recent controversy \cite{Anu,Zurek1}, in terms of the
parameters of the Hamiltonian. Because of the finite relaxation
time of the particle, complete freezing will not take take place,
and the particle will occasionally tunnel from one well to the other.
I think this new way of looking at the dynamics of the tunneling
particle helps in understanding the non-classical behavior seen here.

Coming back to the issue of the dynamics of the spin-boson model,
the value of $ \alpha $ is crucial in deciding whether the
spin can have coherent dynamics or not. The coherent behavior is
destroyed at $ \alpha =1/2$. At $\alpha = 1/2$, $C(t)$ decays with
a single relaxation rate ${\pi\Delta^{2} / 2\omega_{c}}$, which is
what DBGA also yields.

\subsection{The case $1/2\le \alpha <1$}

Beyond $ \alpha =1/2$ we enter the interesting regime
where the spin-boson model is related to the Kondo Hamiltonian.
As mentioned earlier, the results (23)-(25) contain $ \Delta$,
and hence $ J_{\perp }$, to arbitrary order. Hence, the
calculation is expected to be better than that in \cite{tqkondo}.
In order to obtain exponential relaxation rates which are relevant
for experimentalist, I first neglect the frequency dependence
in the arguments of the Gamma functions in (23). The function
$\hat{C}(s)$ is anyway concentrated around $ \omega $ =0. This
results in the following form of the correlation function
\begin{equation}
 \hat{C}(s) \approx { 1\over s + F(0) + [F(\Delta )+F(-\Delta )]/2 +
{[F(\Delta)-F(-\Delta )]^{2 }/4\over s + [F(\Delta )+F(-\Delta )]/2}}.
\end{equation}
The above expression can be decomposed into partial fractions as
\begin{equation}
 \hat{C}(s) = { A\over s + \Gamma_{1}} + { B\over s + \Gamma_{2}} ,
\end{equation}
where
\begin{eqnarray}
 \Gamma_{1,2 }&=& { 1\over 2} [F(0)+F(\Delta )+F(-\Delta)] \nonumber\\
&&\pm {1\over 2}\sqrt{ F^{2 }(0)-[F(\Delta )-F(-\Delta )]^{2}}, \nonumber\\
 A,B &=& { 1\over 2} \pm  { F(0)/2\over \sqrt{ F^{2 }(0)-[F(\Delta)-
F(-\Delta )]^{2}} }
\end{eqnarray}
The spin shows two exponential decay rates $\Gamma_{1}$ and $\Gamma_{2}$.
The Quantum Monte Carlo simulation of the spin-boson model by Egger and
Weiss revealed that after an initial exponential relaxation, the spin shows
a slower exponential relaxation \cite{Egger}. This second slower decay
rate was estimated
to be roughly half the initial decay rate. The initial decay
rate is seen to match the DBGA value. In order to make contact
with the QMC prediction I consider the special case $ F^{2}(0)\gg
-[F(\Delta )-F(-\Delta )]^{2 }$ which is satisfied for the parameters
chosen by Egger and Weiss. For this case I arrive at
\begin{eqnarray}
\Gamma_{1 } &\approx& F(0)+[F(\Delta )+F(-\Delta )]/2 \nonumber\\
\Gamma_{2 } &\approx& [F(\Delta )+F(-\Delta )]/2 \nonumber\\
 A &\approx& 1 - { -[F(\Delta ) - F(-\Delta )]^{2 }/2\over F^{2
}(0)-[F(\Delta )-F(-\Delta )]^{2}} \nonumber\\
 B &\approx& {\mp (\Delta ) - F(-\Delta )]^{2 }/2\over F^{2 }(0)-[F(\Delta
)-F(-\Delta )]^{2}}
\end{eqnarray}
Clearly in (32), $ \Gamma_{2 }\approx \Gamma_{1 }$ /2.
Moreover the quantity $B= -[F(\Delta )-F(-\Delta )]^{2}$ being small,
$ \Gamma_{1 }$ dominates for small times. The second decay being slower,
shows up at long times.

In the Kondo regime, expression (23) for $\hat{C}(s)$ is valid for all
temperatures of practical interest. Thus the dynamics is also described
at very low temperature and small frequency, a regime inaccessible
in \cite{tqkondo}. To stress this point let us look at the absorptive
part of the spin susceptiblity $ \chi ''(\omega)$. An exact relation
derived by Shiba \cite{Shiba} for the general Anderson model says that
at $T=0$, $ \chi ''(\omega )$ should vanish linearly in $ \omega$ :
\begin{equation}
 \lim_{\omega \to 0 }{ \chi ''(\omega )\over \omega} = 2\pi \chi_{0}^{2 }.
\end{equation}
This implies that $C(t)$ will asympotically behave as $\sim t^{-2 }$
\cite{Leggett}. DBGA predicts an asymptotic behavior
$\sim t^{-2(1- \alpha)},$
and hence a divergent spectral function $S(\omega)$. Very recently,
conformal field theory has been employed to calculate the exact
asymptotic spin correlation function for the Kondo Hamiltonian
\cite{Affleck}. The exact results show that the spin correlation
function asymptoticaly decays as $\sim t^{-2}$. Let us take a closer
look at the $ \chi ''(\omega )$ calculated here. In order to make the
complicated expression (25) more tractable, I consider the case
$\omega < k_{B}T/\hbar\ll \Delta$. In this situation $F(\omega_{\pm})$
become temperature independent, and $F(\omega)$ becomes negligible
in their comparison. It is convenient to introduce a parameter
$\epsilon\equiv 1-\alpha$ which is related to the $J_{\parallel}$
in the Kondo problem. In reality, it may be treated as a small
parameter. For small $\omega$ one may expand the various terms
involved in a Taylor series :
\begin{eqnarray}
 F(\omega_{+}) + F(\omega_{-}) &\approx& 2\gamma \Delta^{1-2\epsilon}
\sin(\pi \epsilon ) + 2i\gamma\omega\Delta^{-2\epsilon }\cos(\pi\epsilon)
\nonumber\\
 F(\omega_{+}) - F(\omega_{-}) &\approx& 2i\gamma \Delta^{1-2\epsilon}
\cos(\pi \epsilon )
+ 2\gamma \omega \Delta^{-2\epsilon }\sin(\pi \epsilon )
\end{eqnarray}
where
\begin{equation}
 \gamma \equiv {\Delta^{2}\over 4\omega_{c}} {\pi \over\Gamma
(2\epsilon) \sin(\pi \epsilon )} .
\end{equation}
In this approximation $ \chi''(\omega ) $ assumes
the following form
\begin{eqnarray}
&& \chi ''(\omega ) \approx {1\over\pi}Re\nonumber\\
&&{i \Delta + \omega \tan(\pi \epsilon) \over
[{ i\omega \Delta^{2\epsilon }\over\gamma \cos(\pi \epsilon)}
+ \Delta \tan(\pi \epsilon )+i\omega ]^{2 }-
{\gamma\over \Delta^{2\epsilon}} \cos(\pi \epsilon )[i\Delta
+ \omega \tan(\pi \epsilon )]^{2 }} \nonumber\\
\end{eqnarray}
{}From the above expression it is clear that, $\lim_{\omega\rightarrow 0}
\chi ''(\omega )\sim \omega$. This implies
a finite static susceptiblity, and an asymptotic correlation
function going as $\sim t^{-2 }, $ which is a rigorous
result from conformal field theory \cite{Affleck}. The major
achievement of this work is that the low temperature and long
time dynamics of the spin in the Kondo regime is solved and the
results seem quite satisfactory.

In conclusion, I have presented a calculation of the dynamics
of a strongly damped two-level system which goes beyond the
DBGA. In the coherent regime the results substantially deviate
from the corresponding DBGA ones. In the Kondo regime, the short
and intermediate time behavior agrees with the Quantum Monte
Carlo simulation of Egger and Weiss. The spin crosses over to
a slower exponential relaxation after an initial fast decay.
The asymptotic behavior of the spin is in agreement with the
exact conformal field theory results. An interesting relation
to the quantum Zeno effect is presented. Finally, it should be
mentioned that it is straightforward to carry out a similar
calculation for a biased two-level system.

\end{document}